# Controlled Quantum Operations of a Semiconductor Three-Qubit System

Hai-Ou Li[1,3,*], Gang Cao[1,3,*], Guo-Dong Yu[2,3,*], Ming Xiao[2,3,a)], Guang-Can Guo[1,3], Hong-Wen Jiang[4], and Guo-Ping Guo[1,3,b)]

[1] *Key Laboratory of Quantum Information, CAS, University of Science and Technology of China, Hefei, Anhui 230026, China*
[2] *Department of Optics and Optical Engineering, University of Science and Technology of China, Hefei, Anhui 230026, China*
[3] *Synergetic Innovation Center of Quantum Information & Quantum Physics, University of Science and Technology of China, Hefei, Anhui 230026, China*
[4] *Department of Physics and Astronomy, University of California, Los Angeles, CA 90095, USA*

[*] *These authors contributed equally to this work.*
[a),b)] *Authors to whom correspondence should be addressed. Electronic mail addresses:*
*maaxiao@ustc.edu.cn; gpguo@ustc.edu.cn*

**Abstract**

The Coulomb interactions between electrons play important roles in coupling multiple qubits in various quantum systems. Here we demonstrate controlled quantum operations of three electron charge qubits based on three capacitively coupled semiconductor double quantum dots. The strong interactions between one double dot and other two double dots enable us to control the coherent rotations of one target qubit by the states of two control qubits.

Qubits based on semiconductor quantum dots have made considerable progress in the recent years. Single electron spin qubits [1-6], single charge qubits [7-10], or single spin-charge hybrid qubits [11-12], have been demonstrated in both GaAs and silicon devices. Strongly coupled two-qubit systems, based on the Coulomb interaction between neighboring electrons, have also been realized in either spin [13-16] or charge [17-18] qubit systems. Therefore arbitrary single-qubit gates and two-qubit gates have been realized, which could principally be used to build up any quantum operations.

In this paper we demonstrate a three-qubit system based on the same architecture as earlier semiconductor qubit systems. The realization of more sophisticated quantum logic gates, such as a three-qubit Toffoli gate [19], will make quantum computation more effective, and hence relieving the requirement of quantum error correction when performing multiple quantum logic gates. In addition, it has been predicted that three coupled qubits will be important to study quantum correlation among electrons in semiconductors, such as GHZ states [20, 21].

Our three-qubit system comprised of three capacitively coupled double quantum dots (DQDs). Each DQD behaves as a single charge qubit, based on the control on the location of an individual electron on either side of a double dot with electrical voltage pulses in GHz speed. Strong coupling energies, originating from Coulomb interactions between neighboring electrons charges, arise between different DQDs. This allows us to control the energy level of the electron on one DQD by the locations of the two electrons on other two DQDs. Eventually, we will show how to control the quantum operations of a target qubit, including both its amplitude and phase rotations, by the states of two control qubits.

The device was defined via electron-beam lithography on a molecular-beam-epitaxially grown GaAs/AlGaAs heterostructure. A two-dimensional electron gas (2DEG) is present 95 nm below the surface. The 2DEG has a density of $3.2 \times 10^{11}$ cm$^{-2}$ and a mobility of $1.5 \times 10^5$ cm$^{-2}$V$^{-1}$s$^{-1}$. Fig. 1(a) is a scanning electron microscopy (SEM) image of a typical device. On the top, the first DQD is defined by gates $U_1 - U_5$, $H_L$, and $H_R$, and the second DQD is defined by gates $U_5 - U_9$, $H_L$, and $H_R$. On the bottom, the third DQD is defined by gates $L_1 - L_5$, $H_L$ and $H_R$. Three quantum point contacts (QPCs), defined by gates $Q_1$, $Q_2$, and $Q_3$, respectively, detect the location of a valance electron on each DQD. The experiments are performed in an Oxford Triton dilution refrigerator with a base temperature of 10 mK. Standard lock-in modulation and detection techniques are used for the charge-sensing readout.

Taking each DQD as a charge qubit and considering the inter-qubit coupling, the Hamiltonian of this three-qubit system can be written as follows:

$$H_{3q} = \frac{\varepsilon_1 \sigma_z + \Delta_1 \sigma_x}{2} \otimes I \otimes I + I \otimes \frac{\varepsilon_2 \sigma_z + \Delta_2 \sigma_x}{2} \otimes I + \frac{\varepsilon_3 \sigma_z + \Delta_3 \sigma_x}{2} \otimes I \otimes I$$
$$+ J_{12} \frac{I - \sigma_z}{2} \otimes \frac{I - \sigma_z}{2} \otimes I + J_{13} \frac{I - \sigma_z}{2} \otimes I \otimes \frac{I - \sigma_z}{2} \quad (1)$$

Here, $\varepsilon_i$ and $\Delta_i = 2t_i$ (i = 1, 2, 3) are the energy detuning and twice the inter-dot tunneling rate for each DQD, respectively. $\sigma_x$ and $\sigma_z$ are the Pauli matrixes and $I$ is the identity matrix. $J_{12}$ is the inter-qubit coupling energy between qubit-1 and qubit-2, and $J_{13}$ is that between qubit-1 and qubit-3. Qubit-1 and qubit-2 are closely neighbored and we can deliberately tune voltage on gates $U_5$, $H_L$, and $H_R$ to achieve strong capacitive coupling between them and nearly zero tunnel coupling at the same time. The gap distance between gates $H_L$ and $H_R$ is very narrow (< 80 nm) and this allows us to tune $J_{13}$ into strong regime while forbidding direct tunneling between qubit-1 and qubit-3 [18, 22]. On the contrary, the Coulomb interaction between qubit-2 and qubit-3 is screened by gate $H_R$ and their coupling energy is normally measured to be less than one fifth of $J_{13}$. Therefore, the coupling between qubit-2 and qubit-3 is neglected in our consideration. Here we set the qubit-1 as the target qubit and the other two qubits as the control qubits. We will demonstrate that a three-qubit gate is possible based on strong enough coupling between the target qubit and the two control qubits.

We denote the eight eigenstates of $H_{3q}$ as: |000>, |100>, |010>, |110>, |001>, |101>, |011> and |111>. The definition of |0> and |1> states for each qubit is indicated by the labels in Fig. 1, where $|0>_1|0>_2|0>_3$ corresponds to |000> and so on. For consistency, we define |0>'s as the states that the neighboring qubits are far apart from each other and define |1>'s as the states that the neighboring qubits are closely located. In the later contexts we will describe the evolution of the qubits in the frame of these eigenstates.

Depending on the specific locations of a valance electron in each DQD, the coupling among three-qubits will give rise to different interaction energies. When the electron in the target qubit is far from the two electrons in both control qubits, corresponding to the state |000>, the interaction energy is smallest and we denote it as zero. When the target electron is close to the second electron and far from the third electron, corresponding to state |110>, the interaction energy is $J_{12}$. When the target electron is far from the second electron and close to the third electron, corresponding to state |101>, the interaction energy is $J_{13}$. The largest interaction energy, $J_{12}+ J_{13}$, arises if the target electron is close to both control electrons, corresponding to state |111>.

In Figs. 1(b)-(e), we present the experimentally measured coupling energies $J_{12}$ and $J_{13}$. In Figs. 1(b)-(c), we fix qubit-3 at $|0>_3$ state ($\varepsilon_3 << 0$) and sweep the detuning of qubit-1 against the detuning of qubit-2 (sweep $\varepsilon_1$ against $\varepsilon_2$). The detuning energies $\varepsilon_1$ and $\varepsilon_2$ are converted through the voltage on the relative plunger gates, $U_2$, $U_4$, $U_6$, and $U_8$, using a 35 μeV/mV lever arm. The z-axis is the differential current of the QPC-1 and QPC-2, respectively. We see an abrupt shift in each QPC signal around the anti-crossing points: $\varepsilon_1 = 0$ and $\varepsilon_2 = 0$. As both $\varepsilon_1$ and $\varepsilon_2$ goes from the negative side through the anti-crossing point to the positive side, these two qubits changes from $|0>_1|0>_2$ state to $|1>_1|1>_2$ state and the three-qubit state correspondingly changes from |000> to |110>, as illustrated by the labels in these figures. In |110> state, the coupling energy $J_{12}$ between the first two qubits, as explained above, shifts the anti-crossing point of each qubit towards the positive side. The amount of this abrupt shift directly tells us the value of $J_{12}$, which is about 105 μeV in this case [22-23].

In Figs. 1(d)-(e), we fix qubit-2 at $|0>_2$ state ($\varepsilon_2 << 0$) and do the same measurement for qubit-1 and qubit-3 by sweeping sweep $\varepsilon_1$ against $\varepsilon_3$. In the same way, the abrupt energy shift around the anti-crossing points ($\varepsilon_1 = 0$ and $\varepsilon_3 = 0$) tells the value of their coupling strength: $J_{13}$ approximately equals 135 μeV.

These large coupling strengths will enable us to implement three-qubit operations [18, 24]. For instance, the coherent rotations of the target qubit, such as Larmor procession [8], Landau-Zener-Stütckelberg (LZS) interferences [9], and Rabi oscillations [10], usually occur when the qubit is brought from an initial state to its detuning anti-crossing point. Because the target qubit's anti-crossing point shifts when coupled with the control qubits, we can control its coherent rotations by switching the states of the control qubits, or intuitively, by moving the electrons on the control qubits towards to or away from the electron on the target qubit. And if the inter-qubit coupling is strong enough, the shift in the target qubit's anti-crossing point will be large enough, and we will be able to completely switch on/off the target qubit's coherent rotations, and therefore achieve high-fidelity three-qubit quantum gates.

The mechanism of controlled coherent rotations of the target qubit is illustrated in Figs. 2(a)-(e). Taking the example of the Larmor procession, we initialize the target qubit at $|0>_1$ state ($\varepsilon_1 << 0$) and apply a non-adiabatic voltage pulse to drive it to its anti-crossing point ($\varepsilon_1 = 0$) where coherent rotations between $|0>_1$ and $|1>_1$ states occur. In fact, the same mechanism applies in the case of LZS interferences in which an adiabatic passage is applied to the target qubit to bring it to its anti-crossing point where a non-adiabatic transition occurs, or Rabi oscillations in which a microwave pulse is applied and the coherent rotations occur at the anti-crossing point where the resonance condition is met.

In Fig. 2(a), we depict the simulated energy levels of all the eight eigenstates for the three-qubit Hamiltonian $H_{3q}$. We fixed the states of the control qubits and calculated the variation of the energy spectroscopy with respect to the detuning of the target qubit. In this figure $\varepsilon_2 = -200$ μeV and $\varepsilon_3 = -200$ μeV so that the ground state of the control qubits is $|0>_2|0>_3$. The values of $J_{12}$ and $J_{13}$ are chosen as the experimental obtained values. For guidance, we labeled the states corresponding to each energy level, in places where $\varepsilon_1 << 0$ and $>> 0$. In particular, we are interested in the ground stats and the first excited state of the target qubit. In the energy range within which we are doing operations, to say, -200 μeV < $\varepsilon_1$ < 200 μeV, the two lowest energy states are $|000>$ and $|100>$, as shown by the blue solid and dashed lines in Fig. 2(a).

In Figs. 2(b)-(e), we varied the ground states of the control qubits and focus on the few lowest energy states. Fig. 2(b) is a magnification of Fig. 2(a), in which $\varepsilon_2$ and $\varepsilon_3$ are fixed at both -200 μeV. As expected, the inter-qubit coupling is weak and the anti-crossing point between states $|000>$ and $|100>$ remains at $\varepsilon_1 = 0$, just like the case of a single qubit. If we initialize the target qubit at $\varepsilon_1 << 0$, for instance, at -150 μeV, and pulse it to $\varepsilon_1 = 0$, coherent rotations between $|000>$ and $|100>$ will occur.

In Fig. 2(c), the values of $\varepsilon_2$ and $\varepsilon_3$ are set as 200 and -200 μeV and the ground states of the control qubits is now $|1>_2|0>_3$. In this condition, we see that the anti-crossing point between the two lowest $|0>_1$ and $|1>_1$ states, which are labeled as states $|010>$ and $|110>$ by the blue solid and dashed lines, shifts to $\varepsilon_1 = J_{12} = 105$ μeV. This is due to the coupling energy between qubit-1 and qubit-2, as explained above. Consequently, if we still initialize the target qubit at $\varepsilon_1 = -150$ μeV and pulse it to $\varepsilon_1 =$ 0 μeV, the target qubit is now still below the anti-crossing point, and the coherent rotations are suppressed. After the pulse finishes, the three-qubit state will simply return to the initial state |010>.

The suppression of coherent oscillations lies in the following facts. First of all, the qubit coherence time exponentially decreases as a dependence on the detuning away from the anti-crossing point [8]. Therefore the anti-crossing point is usually the "sweet spot" for coherent rotations whereas below the anti-crossing point the rotations diphase too quickly to persist. Even without decoherence effect, for this two-level anti-crossing quantum system, the rotation amplitude of Larmor procession, LZS interferences, and microwave-driven Rabi oscillations reaches the maximum at the anti-crossing point and deteriorates quickly with respect to the detuning.

In Fig. 2(d), the values of $\varepsilon_2$ and $\varepsilon_3$ are set as -200 and 200 μeV and the ground states of the control qubits changes to $|0>_2|1>_3$. Due to the coupling energy between qubit-1 and qubit-3, the anti-crossing point between |001> and |101> states shifts to $\varepsilon_1 = J_{13} = 135$ μeV. Finally, in Fig. 2(e) the values of $\varepsilon_2$ and $\varepsilon_3$ are set as both 200 μeV and the ground states of the control qubits changes to $|1>_2|1>_3$. The anti-crossing point between |011> and |111> states shifts to $\varepsilon_1 = J_{12} + J_{13} = 240$ μeV. Like the case of Fig. 2(c), in these two cases the coherent rotations will both be suppressed due to the tremendous energy shift of the target qubits's anti-crossing point. As a conclusion, coherent rotations of the target qubit that occur when both control qubits are in |0> states will be suppressed if either control qubit changes to |1> state.

Now we introduce our experimental realization of controlled coherent rotations for this three-qubit system. We prepare the three qubits at an initial state |000> and apply a non-adiabatic rectangular voltage pulse to the target qubit. In this condition, we choose the pulse amplitude to drive the target qubit exactly to the anti-crossing point between state |000> and |100>. Coherent Larmor procession will occur. The target's amplitude is going to rotate between $|0>_1$ and $|1>_1$ states, by an angle determined by $2\pi\Delta_1 W_1$, where $W_1$ is the pulse width. This allows us to coherently manipulate the amplitude of the target qubit's wave-function.

If we vary the states of the control qubits so that the three-qubit's initial state changes to |010>, |001>, or |011>, the anti-crossing point between the lowest $|0>_1$ and $|1>_1$ states shifts by $J_{12}$, $J_{13}$, or $J_{12}+J_{13}$, respectively. In any case, the energy shift is so large that the same rectangular voltage pulse drives the target qubit far below its anti-crossing point. The Larmor procession is suppressed and the target qubit's amplitude remains unchanged. If the pulse width is chosen so that $2\pi\Delta_1 W_1 = \pi$, a three-qubit Toffoli gate can be realized: the target's state is flipped if and only if the two control qubits are in $|0>_2|0>_3$ state.

The experimental results are illustrated in Figs. 3(a)-(b). The differential current detected by the QPC of the target qubit is shown in both figures. In Fig. 3(a) we fix qubit-2 at $|0>_2$ state and in Fig. 3(b) we fix qubit-2 at $|1>_2$ state. In both figures we scan detuning $\varepsilon_3$, as illustrated by the y-axis, to vary qubit-3 from state $|0>_3$ to $|1>_3$. Therefore, we cover all four eigenstates of the two control qubits: $|0>_2|0>_3$, $|0>_2|1>_3$, $|1>_2|0>_3$, and $|1>_2|1>_3$, as labeled in Figs. 3(a)-(b). The x-axis in these figures is the pulse width on the target qubit, $W_1$. Clearly, coherent oscillations of the target qubit is observed for configuration $|0\rangle_2|0\rangle_3$. This is the Larmor procession, with a frequency $\Delta_1 = 4.5$ GHz, consistent to the value obtained from photon-assisted-tunneling (PAT) measurements [25]. In all other three conditions, no visible Larmor procession is observed. Fig. 3(c) shows representative curves for each of these four conditions, after refinement scanning and many times averaging. The black curve shows a typical Larmor procession trace when the control qubits are in state $|0\rangle_2|0\rangle_3$, while the other three curves show just background fluctuations when the control qubits are in states $|0\rangle_2|1\rangle_3$, $|1\rangle_2|0\rangle_3$, or $|1\rangle_2|1\rangle_3$. These results clearly demonstrate that we can control the target qubit's amplitude rotations by the two control qubits' states.

In Figs. 3(d)-(e), we give simulation to Figs. 3(a)-(b) by numerically solving the Liouville-von Neumann equation regarding the three-qubit Hamiltonian described by Equa. (1). We use relevant parameters obtained from experimental data, including decoherence time $T_2^* = 1.2$ ns. The simulations agree with the experimental results: coherent oscillations are suppressed if either one control qubit is in $|1\rangle$ state. We also noticed that in the region around the balance point, i.e, $\varepsilon_1 \sim 0$, there exist some complex oscillations, with varying frequency and prominent amplitude. These are due to the entanglement between different qubits in the neighborhood of the balance point [26]. This feature is seen in our experimental data in Fig. 3(a), which contains a dark region around $\varepsilon_1 \sim 0$. However, details in this region are not clear enough to extract information about qubit entanglement, possibly because the non-adiabacity of the voltage pulse wipes out the high-frequency component in the complex rotation patterns of entangled qubits [26].

Apart from controlling the amplitude rotations of the target qubit, we can also control its phase rotations by the states of the two control qubits. As we noticed in earlier experiments [9], when we apply an ultrafast voltage pulse (100 to 350 ps in width) on the target qubit's gate, the transmission line acts as a low-pass filter and effectively modifies the pulse into a Gaussian-like shape. As a result, the pulse behaves adiabatically when the pulse width is ultra-short, and we observe LZS interferences [9]. LZS interferences come from a non-adiabatic transition at the anti-crossing point. The non-adiabatic transition corresponds to a rotation of the wave function's amplitude, and the adiabatic evolution following the Gaussian-like voltage pulse gives rise to phase rotations. The frequency of amplitude rotations is determined by target qubit's inter-dot tunneling rate $\Delta_1$, which is about 4.5 GHz in this experiment. The frequency of phase rotations is determined by both the height and width of the adiabatic voltage pulse, and can be tuned much faster than that of the amplitude rotations [9]. In this experiment, it reaches about 15 GHz. Therefore, in the LZS pattern we mainly focus on the phase rotations of the studied qubit.

The strong inter-qubit coupling enables us to control the phase rations of the target qubit. First, we initialize the three qubits at an initial state $|000\rangle$. We apply to the target qubit an ultra-short voltage pulse, which can be regarded as a Gaussian-like pulse. We choose the pulse height such that it drives the target qubit across the anti-crossing point and induces LZS interferences. In this condition, we will see fast phase rotations of the target qubit. Next, we vary the three-qubit state to $|010\rangle$, $|001\rangle$, and $|011\rangle$, and the anti-crossing point between the lowest $|0\rangle_1$ and $|1\rangle_1$ states will shift by $J_{12}$, $J_{13}$, and $J_{12} + J_{13}$, respectively. In any of these three conditions, the pulse will be unable to drive the target qubit above its anti-crossing point any more, and phase rotations will cease.

Figs. 4(a)-(b) shows the experimentally measured data. The x-axis of all figures is the width of the voltage pulse on the target qubit-1, $W_1$, which is within the range of 100 ps to 350 ps. The y-axis of both figures is detuning $\varepsilon_3$, by scanning which we change qubit-3 from $|0>_3$ state to $|1>_3$ state. Qubit-2 is set at $|0>_2$ state in Fig. 4(a) and at $|1>_2$ state in Fig. 4(b). Therefore, we are comparing the phase rotations of the target qubit for all four configurations of the two control qubits' states: $|0>_2|0>_3$, $|0>_2|1>_3$, $|1>_2|0>_3$, and $|1>_2|1>_3$.

For $|0>_2|0>_3$, Fig. 4(a) shows persistent fast rotations. The rotation frequency slightly varies with pulse width due to the non-linear dynamics of LZS interferences [9, 27]. Typically, we can see that a $2\pi$ phase rotation costs about 65 ps, corresponding to a frequency of about 15 GHz. This is indeed much faster than the amplitude rotation frequency. For configurations $|0>_2|1>_3$ and $|1>_2|0>_3$, these fast and clear rotations are gone. There are some residual background fluctuations. These are because of the large pulse amplitude used to drive the target qubit across, more than exactly to, its anti-crossing point to induce LZS interferences for the $|0>_2|0>_3$ configuration. For the $|0>_2|1>_3$ and $|1>_2|0>_3$ configurations, the pulse may reach close to the anti-crossing point and therefore causes residual oscillations. Nonetheless, both experimental data and theoretical simulations show that the residual oscillations are very weak compared with that for the $|0>_2|0>_3$ configuration. Increasing coupling strengths further more will completely eliminate these residual effects. For configuration $|1>_2|1>_3$, the shift of anti-crossing point nearly doubles those of the former two configurations, In this case, the pulse falls a lot more below the anti-crossing point and the residual fluctuations disappear, meaning that the phase rotations are completely suppressed.

The simulations in Figs. 4(c)-(d) support our experimental results: fast and strong phase rotations of the target qubit are allowed when two control qubits are both in state $|0>$. The phase rotations are largely suppressed when one control qubit turns into state $|1>$, and are completely turned off when both control qubits are in state $|1>$.

We have shown that we can control both the coherent amplitude and phase rotations in a three-qubit system. In principle, a non-adiabatic pulse and an adiabatic pulse can be combined to make arbitrary quantum operations for a qubit [28]. This indicates the possibility of performing controlled arbitrary quantum operations in a three-qubit system, based on strong inter-qubit coupling strengths.

In summary, in a semiconductor quantum dot based three-qubit system, we have demonstrated gate-voltage-control of strong inter-qubit couplings by using a specially-designed device structure. We successfully demonstrate coherent control of both amplitude and phase of one target qubit by the pre-prepared states of two control qubits and the basic functionalities of Toffoli gate. We hope our first work in semiconductor to go beyond the two-qubit limit will provide a useful insight to the research on multiple qubit systems in semiconductor devices. A natural extension of this work will be the dynamic control of the target qubit's amplitude and phase, to realize the full quantum logical operations of the three qubits.

This work was supported by the National Key R & D Program (Grant No.2016YFA0301700), the Strategic Priority Research Program of the CAS (Grant No. XDB01030000), the National Natural Science Foundation (Grant Nos. 11575172, 61306150, 11304301, and 91421303), and the Fundamental Research Fund for the Central Univesities.

**References**

1. J. R. Petta, et al. Coherent Manipulation of Coupled Electron Spins in Semiconductor Quantum Dots. Science 309, 2180 (2005).
2. H. L. Koppens, et al. Driven coherent oscillations of a single electron spin in a quantum dot Nature 442, 766 (2006).
3. B. M. Maune, et al. Coherent singlet-triplet oscillations in a silicon-based double quantum dot. Nature 481, 344-347 (2012).
4. E. Kawakami, et al. Electrical control of a long-lived spin qubit in a Si/SiGe quantum dot. Nature Nanotech. 9, 666-670 (2014).
5. X. Wu, et al. Two-axis control of a singlet–triplet qubit with an integrated micromagnet. PNAS 111, 11938-11942 (2014).
6. M. Veldhorst, et al. An addressable quantum dot qubit with fault-tolerant control-fidelity. Nature Nanotech. 9, 981 (2014).
7. T. Hayashi, et al. Coherent manipulation of electronic states in a double quantum dot. Phys. Rev. Lett. 91, 226804 (2003).
8. K. D. Petersson, et al. Quantum coherence in a one-electron semiconductor charge qubit. Phys. Rev. Lett. 105, 246804 (2010).
9. G. Cao, et al. Ultrafast universal quantum control of a quantum-dot charge qubit using Landau-Zener-Stuckelberg interference. Nature Commun. 4, 1401 (2013).
10. D. Kim, et al. Microwave-driven coherent operation of a semiconductor quantum dot charge qubit. Nature Nanotech. 10, 243 (2014)
11. D. Kim, et al. Quantum control and process tomography of a semiconductor quantum dot hybrid qubit, Nature 511, 70-74 (2014).
12. G. Cao, et al. A Tunable Hybrid Qubit in a GaAs Double Quantum Dot. Phys. Rev. Lett. 116, 086801 (2016).
13. R. Brunner, et al. Two-qubit gate of combined single-spin rotation and interdot spin exchange in a double quantum dot. Phys. Rev. Lett. 107, 146801 (2011).
14. K. C. Nowack, et al. Single-shot correlations and two-qubit gate of solid-state spins. Science 333, 1269-1272 (2011).
15. M. D. Shulman, et al. Demonstration of entanglement of electrostatically coupled singlet-triplet qubits. Science 336, 202-205 (2012).
16. M. Veldhorst, et al. A two-qubit logic gate in silicon. Nature 526, 410-414 (2015).
17. G. Shinkai, et al. Correlated coherent oscillations in coupled semiconductor charge qubits. Phys. Rev. Lett. 103, 056802 (2009).
18. H. O. Li, et al. Conditional Rotation of Two Strongly Coupled Semiconductor Charge Qubits. Nature Comm. 6, 7681 (2015).
19. A. Fedorov, et al. Implementation of a Toffoli Gate with Superconducting Circuits. Nature 481, 170-172 (2013).
20. M. Neeley, et al. Generation of three-qubit entangled states using superconducting phase qubits. Nature 467, 570 (2010).
21. L. DiCarlo , et al. Preparation and measurement of three-qubit entanglement in a superconducting circuit. Nature 467, 574 (2010).
22. G. D. Yu, et al. Tunable capacitive coupling between two semiconductor charge qubits. Nanotechnology 27, 324003 (2016).

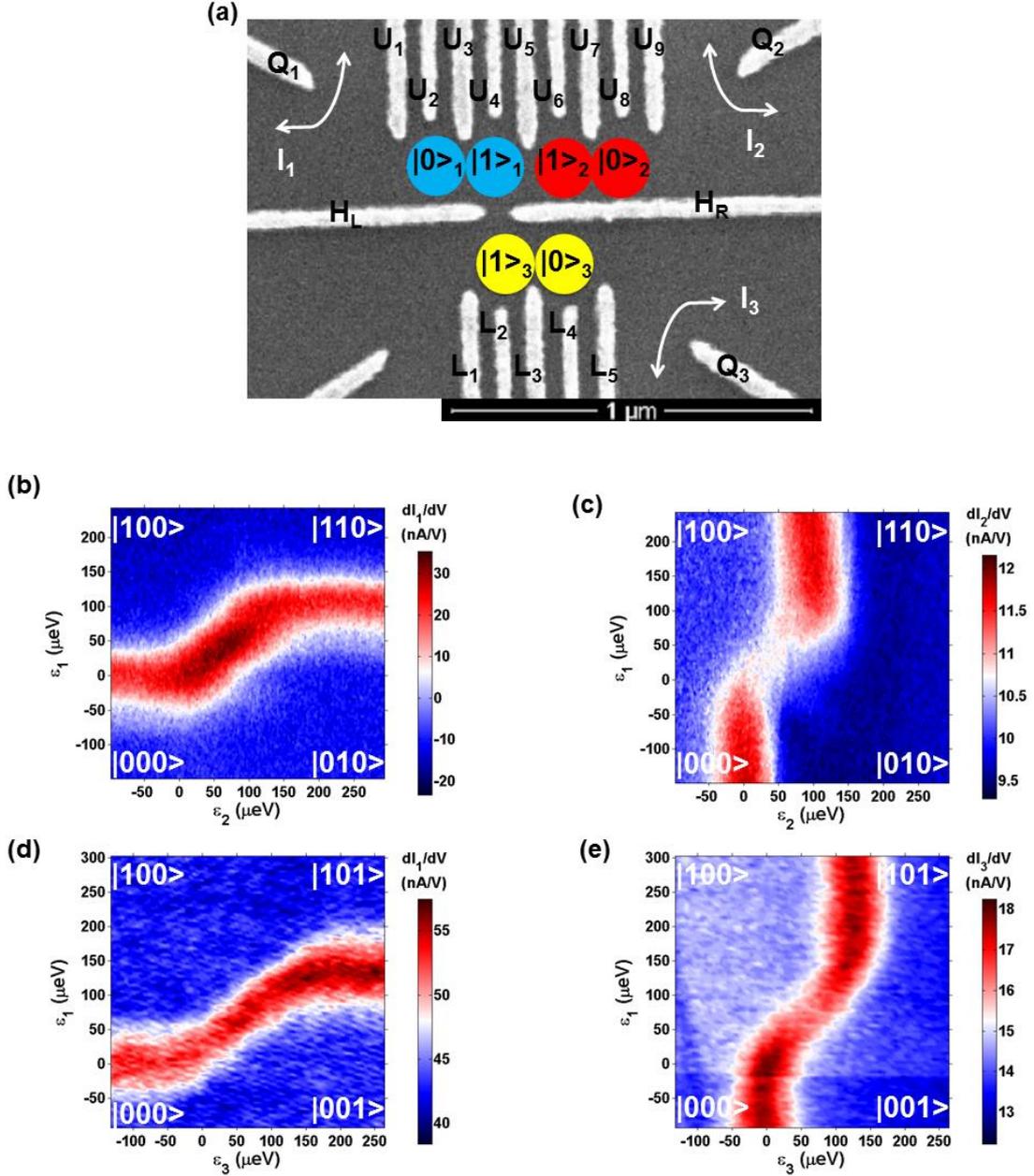

**Figure 1 (a)** SEM image of three coupled qubits. Qubit-1 is defined by gates $U_1 - U_5$, $H_L$ and $H_R$. Qubit-2 is defined by gates $U_5 - U_9$, $H_L$ and $H_R$. Qubit-3 is defined by gates $L_1 - L_5$, $H_L$ and $H_R$. For each qubit, the definition of |0> and |1> states are as labeled. The three QPCs, defined by gates $Q_1$, $Q_2$, and $Q_3$, respectively, detect the location of a valance electron on each DQD. **(b) and (c)** Coupling energy $J_{12}$ between qubit-1 and qubit-2, as manifested by the abrupt shift from state |000> to |110>. Qubit-3 is fixed at $|0>_3$. Represented in these two figures are the differential current measured by QPC-1 and QPC-2, respectively. The x- and y-axis in these two and the following two figures are the detuning of the three qubits, converted from the voltages on corresponding plunger gates. **(d) and (e)** Coupling energy $J_{13}$ between qubit-1 and qubit-3, as manifested by the abrupt shift from state |000> to |101>. Qubit-2 is fixed at $|0>_2$. Represented in these two figures are the differential current measured by QPC-1 and QPC-3, respectively.

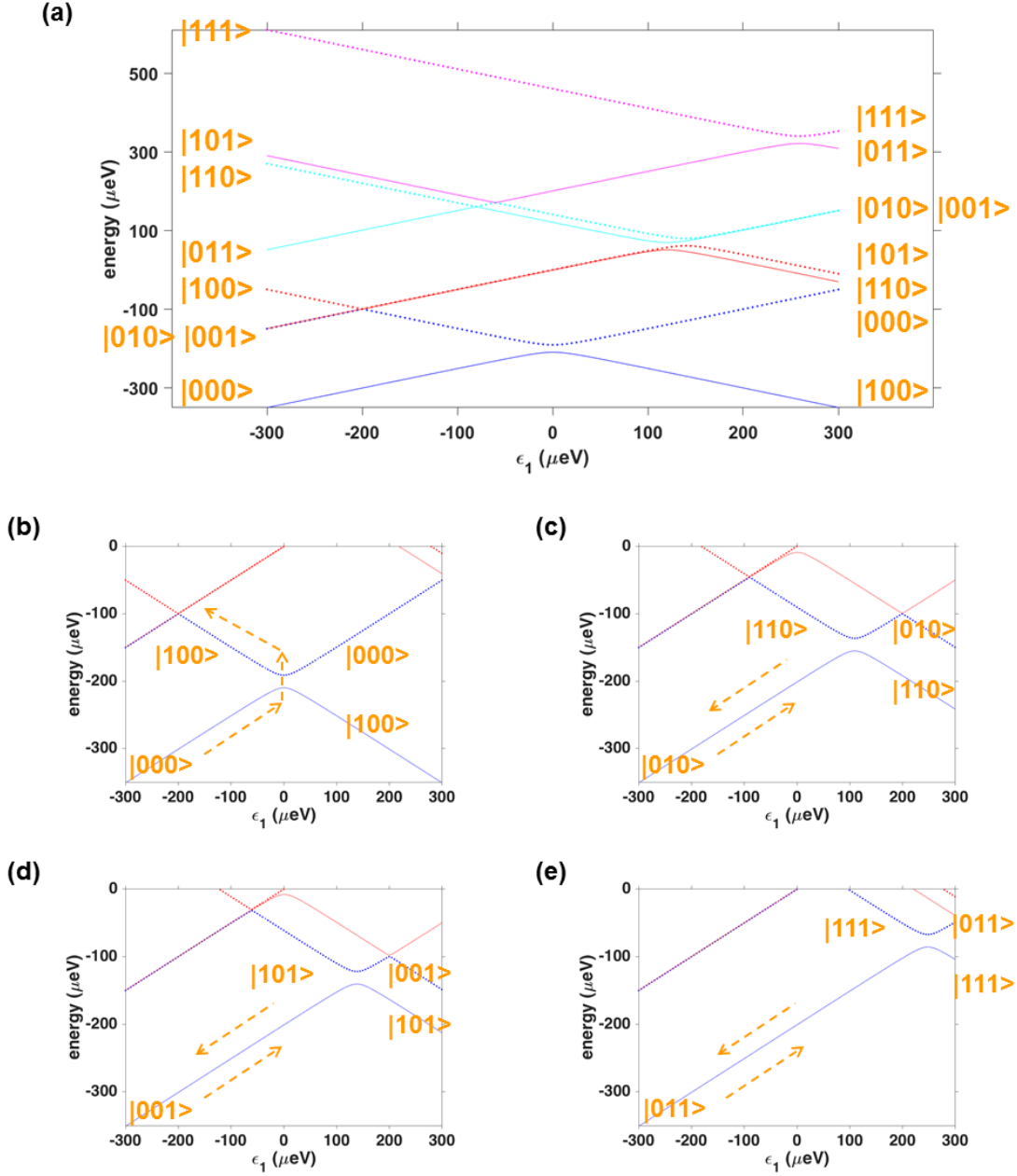

**Figure 2 (a)** Simulated three-qubit energy levels. The detuning of qubit-2 and qubit-3 are fixed far below anti-crossing point. Scanned in the x-axis is the detuning of qubit-1. **(b) – (e)** Illustration of coherent rotations of qubit-1, as controlled by the states of qubit-2 and qubit-3. The detuning of qubit-2 and qubit-3 are such configured so that their ground states are $|0\rangle_2|0\rangle_3$, $|1\rangle_2|0\rangle_3$, $|0\rangle_2|1\rangle_3$, and $|1\rangle_2|1\rangle_3$, respectively. Only the few relevant lowest energy levels are shown in these four figures. In (b), a pulse is designed to drive qubit-1 from $|0\rangle_1$ state to its anti-crossing point when control qubits are fixed at state $|0\rangle_2|0\rangle_3$. Transition occurs at the anti-crossing point and coherent rotations will be initiated. In (c)–(e), the anti-crossing point of qubit-1 shifts to positive side by $J_{12}$, $J_{13}$, and $J_{12} + J_{13}$, respectively. Qubit-1 cannot be driven to its anti-crossing point anymore and it adiabatically evolves back to the initial state.

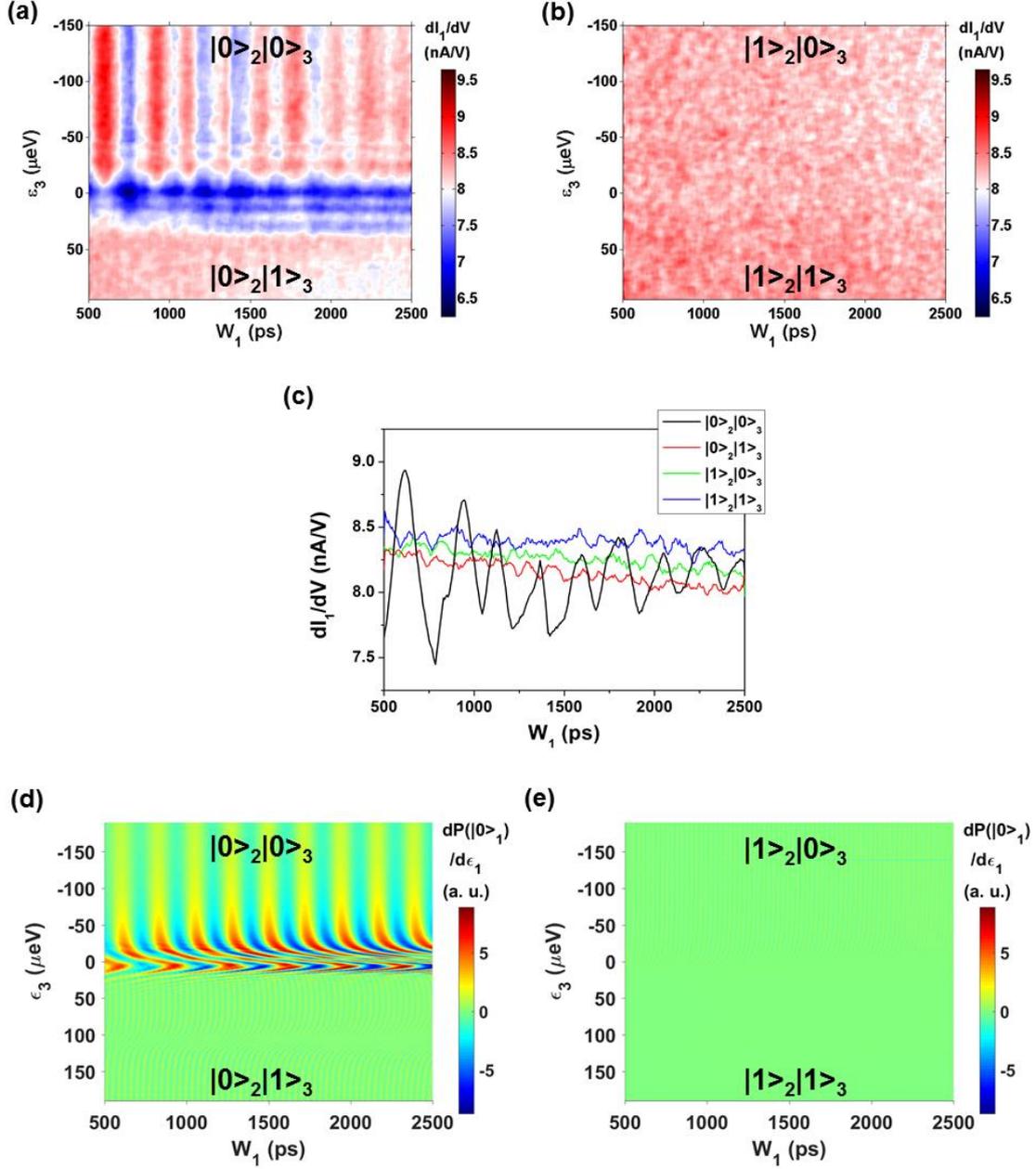

**Figure 3** **(a)** and **(b)** The coherent amplitude rotations of qubit-1, as controlled by the states of qubit-2 and qubit-3. The x-axis, $W_1$, is the width of a non-adiabatic rectangular pulse on the detuning of qubit-1. The y-axis is the detuning of qubit-3, converted from the voltages on plunger gates $V_{L2}$ and $V_{L4}$. Qubit-2 is fixed at state $|0\rangle_2$ and $|1\rangle_2$, respectively. Coherent amplitude rotations at a frequency of around 4.5 GHz are observed only in the first case. For comparison, the z-axis plot ranges of these two figures are set equal. **(c)** Typical traces cut along the x-axis in the above two figures. Qubit-2 and quabit-3 are set as $|0\rangle_2|0\rangle_3$, $|0\rangle_2|1\rangle_3$, $|1\rangle_2|0\rangle_3$, and $|1\rangle_2|1\rangle_3$ for the black, red, green, and blues curves, respectively. **(d)** and **(e)** Theoretical simulations of the differentiated probability on $|0\rangle_1$ state with respect to the detuning of qubit-1, resembling the experimentally measured differential current of QPC-1. Their z-axis plot ranges are intentionally set equal.

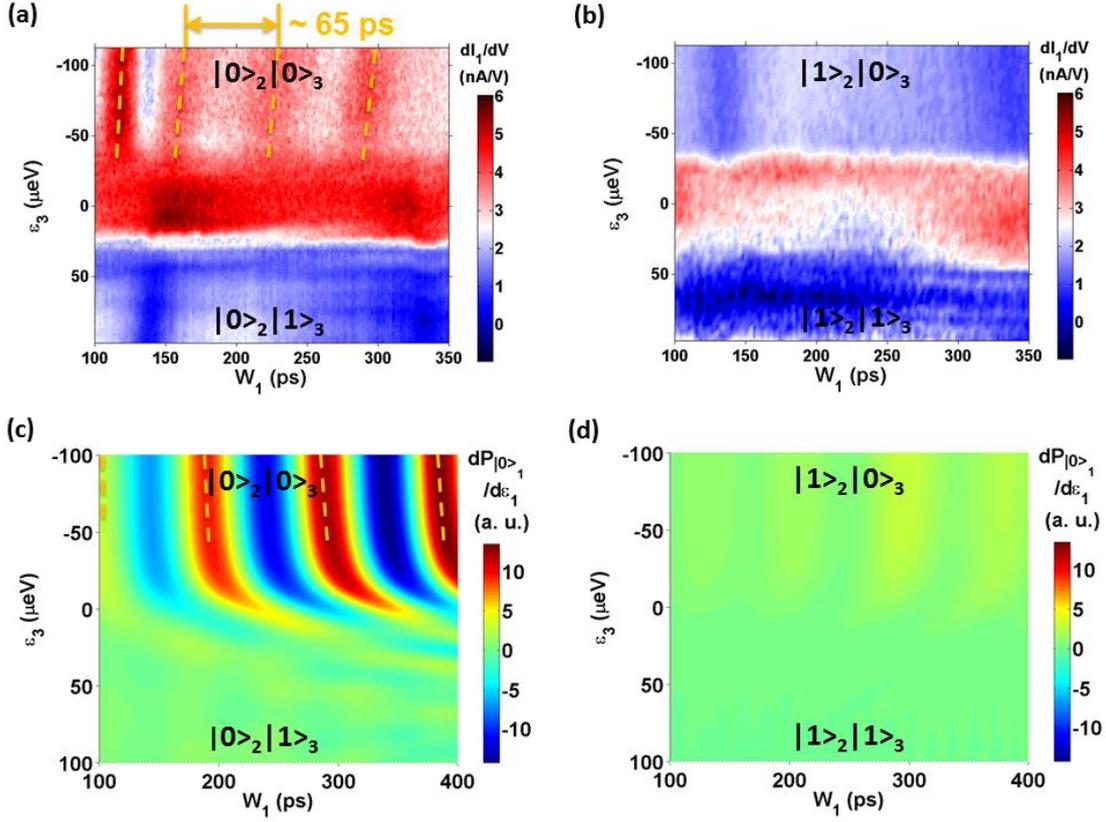

**Figure 4 (a)** and **(b)** The coherent phase rotations of qubit-1, as controlled by the states of qubit-2 and qubit-3. The x-axis, $W_1$, is the width of an adiabatic Gaussian-like pulse on the detuning of qubit-1. The y-axis is the detuning of qubit-3, converted from the voltages on plunger gates $V_{L2}$ and $V_{L4}$. Qubit-2 is fixed at state $|0>_2$ and $|1>_2$, respectively. Fast phase rotations at a frequency of approximately 15 GHz are observed only in the first case. For comparison, the z-axis plot ranges of these two figures are set equal. **(c)** and **(d)** Theoretical simulations of the differentiated probability on $|0>_1$ state with respect to the detuning of qubit-1, resembling the experimentally measured differential current of QPC-1. Their z-axis plot ranges are intentionally set equal.